\documentclass[prl,amsmath,amssymb,twocolumn]{revtex4}
\usepackage{graphics}
\usepackage{epsfig}
\usepackage{bbm}
\begin{document}
\newcommand{\ket}[1]{|{#1}\rangle}
\newcommand{\bra}[1]{\langle{#1}|}
\newcommand{\braket}[1]{\langle{#1}\rangle}
\newcommand{\ad}{a^\dagger}
\newcommand{\e}{\ensuremath{\mathrm{e}}}
\newcommand{\norm}[1]{\ensuremath{| #1 |}}
\newcommand{\aver}[1]{\ensuremath{\big<#1 \big>}}
\title{Spatial correlations of trapped 1D bosons in an optical lattice}

\author{C. Kollath}
\affiliation{Ludwig-Maximilians-Universit\"at, Theresienstr.\ 37, D-80333 M\"unchen, Germany.}
\author{U. Schollw\"ock}
\affiliation{Ludwig-Maximilians-Universit\"at, Theresienstr.\ 37, D-80333 M\"unchen, Germany.}
\author{J. von Delft}
\affiliation{Ludwig-Maximilians-Universit\"at, Theresienstr.\ 37, D-80333 M\"unchen, Germany.}
\author{W. Zwerger}
\altaffiliation[present address: ]{Institute for Theoretical Physics, University of Innsbruck, Technikerstr. 25, A-6020 Innsbruck}%Lines break automatically or can be forced with \\
\affiliation{Ludwig-Maximilians-Universit\"at, Theresienstr.\ 37, D-80333 M\"unchen, Germany.}
\date{\today}

\begin{abstract}

 We investigate a quasi-one dimensional system of trapped cold bosonic atoms in an optical lattice by using the density matrix renormalization group to study the Bose-Hubbard model at $T=0$ for experimentally realistic numbers of lattice sites. It is shown that a properly rescaled one-particle density matrix characterizes superfluid versus insulating states just as in the homogeneous system. For typical parabolic traps we also confirm the widely used local density approach for describing correlations in the limit of weak interaction. Finally, we note that the superfluid to Mott-insulating transition is seen most directly in the half width of the interference peak.
\end{abstract}

%% 05.30.Jp Boson systems
%% 03.75 Lm Tunneling, Josephson effect, Bose-Einstein condensates in
%%          periodic potentials, solitons, vortices and topological excitations
%% 73.43.Nq Quantum phase transitions

\maketitle

% ============================================================

 During the last years enormous progress was made in the experimental manipulation of cold atoms in optical lattices. Recently, Greiner et al. \cite{GreinerBloch2002} succeeded in driving a transition between a superfluid (SF) and a Mott-insulating (MI) state in a system of ultracold bosonic atoms in an optical lattice as predicted by Jaksch et al. \cite{JakschZoller1998}. In contrast to solid state realizations the experimental setup involves the application of an additional parabolic trapping potential that 
causes a state in which the two phases, though spatially separated, coexist \cite{BatrouniTroyer2002}.
Due to the inhomogeneity the usual characterization of the SF to MI transition by the asymptotic behaviour of the one-particle density matrix does not apply. Motivated by this, we use the density matrix renormalization group (DMRG) \cite{White92} to study how the parabolic confining potential influences the one-particle density matrix and its Fourier transform, which is related to the interference pattern observed in the experiments \cite{KashurnikovSvistunov2002}. 
We find that by a simple rescaling, the decay of the correlations can be used to characterize the occuring states, just as in the homogeneous case. We further confirm the applicability of the standard local density approximation to the inhomogeneous system \cite{GangardtShlyapnikov2003} for weak interactions by comparing it to the DMRG results for the correlation functions.
Studying experimentally accessible quantities we find that the half width of the interference peak contains the essential information about the state of the system.

\paragraph{Model:} Ultracold bosonic atoms in an optical lattice \cite{JakschZoller1998} can be described by a Bose-Hubbard model 
\begin{equation}
\label{eq:bh}
H= -J \sum_{j} (b_j^\dagger b^{\phantom{\dagger}}_{j+1}+h.c.) + \frac{U}{2} \sum _j \hat{n}_j ( \hat{n}_j-1)+ \sum_j \varepsilon_j \hat{n}_j, 
\end{equation}
where $b^\dagger_j$ and $b_j$ are the creation and annihilation operators on site $j$ and $ \hat{n}_j= b^\dagger_j b^{\phantom{\dagger}}_j$ is the number operator \cite{FisherFisher1989}. This Hamiltonian describes the interplay between the kinetic energy due to the next-neighbour hopping with amplitude $J$ and the repulsive onsite interaction $U$ of the atoms.
By tuning the lattice depth in the experiment, the parameter $u=U/J$ can be varied over several orders of magnitude.
To investigate the properties of the 1D Bose-Hubbard model, we apply the DMRG, a quasi-exact numerical method, very well suited to study strongly correlated quasi 1D quantum systems with a large number of sites at zero temperature \cite{PeschelHallberg1998}. It has been successfully applied to spin, fermionic and bosonic quantum systems including the homogeneous \cite{KuehnerMonien2000} and the disordered \cite{RapschZwerger1999} Bose-Hubbard model. 
We used the finite size DMRG algorithm \cite{PeschelHallberg1998} which is better suited for an inhomogeneous system, since it gives the system the possibility to evolve further after the final length of the system is reached.
Additionally some tricks
 are applied to circumvent problems which arise due to the sparse filling at the boundaries. The numerical results were tested to be convergent in the cut-offs used for the length of the system, the number of states kept for the Hilbert space, and the number of states allowed per site. Uncertainties given below are determined by comparing data of different parameter sets.

% ============================================================

\begin{figure} 
\begin{center}
  {\epsfig{figure=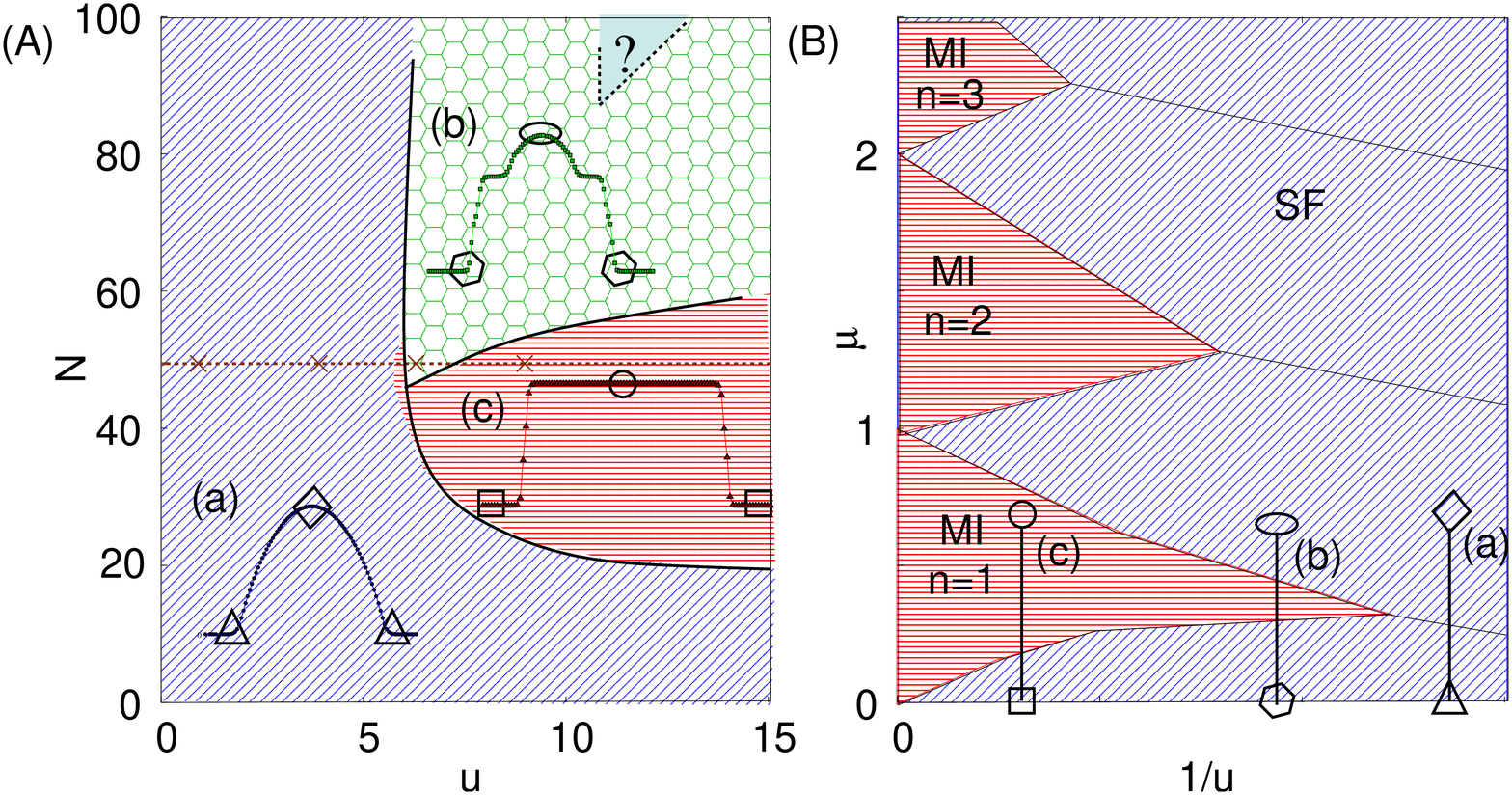,width=0.8\linewidth}}
\end{center}
\caption{A) Sketch of the state diagram for $v_0=4/64^2$. The insets sketch the shape of the density distribution in the states. (B) Sketch of the phase diagram of the homogeneous system: chemical potential $\mu$ versus $1/u$. The different symbols in (B) mark the locations of the chemical potential values in the local density approximation that correspond to the locations in the density profiles marked in (A). 
}
\label{fig:statediasketch}
\end{figure}

\paragraph{State diagram:}
The confining trap of the experiment \cite{GreinerBloch2002} which consists of a magnetic trap and the confining component of the laser which generates the optical lattice, can be modeled by setting $\varepsilon_j= V^0_\textrm{trap}\left(a(j-j_0)\right)^2$ in Eq.\ (\ref{eq:bh}), where $a$ is the lattice constant. We choose the strength of the trap proportional to the onsite interaction, i.e. $V_\textrm{trap}^0 =v_0 U$, since this guarantees that  when the optical lattice depth, corresponding to the parameter $u$ in the Bose-Hubbard model, is changed, the size of the system does not vary much for a fixed particle number.
This is consistent with the experimental realization, in which the total size of the condensate is essentially independent of the lattice depth. In the presence of a parabolic trap at average filling of approximately one particle per site, one can distinguish three states of the system (see \cite{BatrouniTroyer2002,JakschZoller1998}): (a) for $u<u_{c1}$, the particle occupancy is incommensurate over the whole system; (b) for $u_{c1}<u<u_{c2}$, regions with incommensurate and commensurate occupancy coexist; and (c) for $u>u_{c2}$, the main part of the system is locked to commensurate filling and only at the boundaries small incommensurate regions exist. For small particle numbers, state (b) does not occur. A sketch of the state diagram is presented in Fig. \ref{fig:statediasketch} (A). The insets show the characteristic shape of the particle distribution for the three states.
For state (b) the exact locations of the interface between the commensurate and the incommensurate regions are difficult to determine. This is due to the fact that these sites correspond in the homogeneous system to the critical parameter regimes at the phase transition, where strong fluctuations and extreme sensitivity to boundary conditions make a numerical investigation very difficult. 

\begin{figure} 
\begin{center}
        {\epsfig{figure=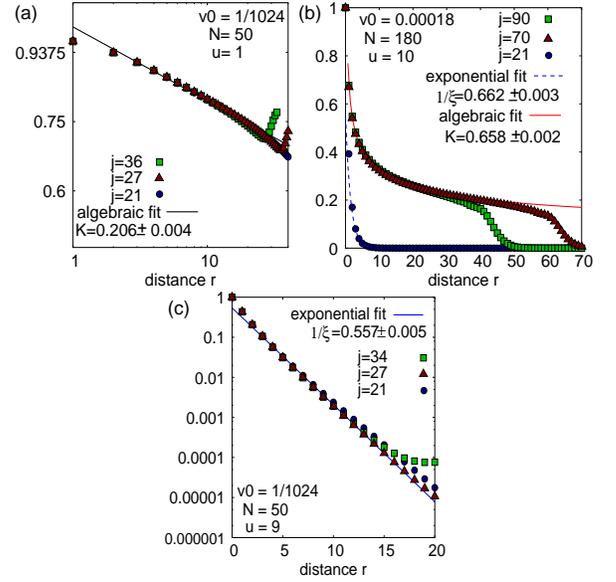,width=0.9\linewidth}}
\end{center}
\caption{Scaled correlations $C_j(r)$ [Eq.\ \ref{eq:corrscal}] for different fixed sites $j$ are plotted as a function of $r$ for different values of $u$. For the coexistence region (b) a shallower trapping potential is chosen, such that the extents of both the incommensurate and the commensurate region are large enough to allow identification of the algebraic and exponential behaviour.
}
\label{fig:corrdecaycomb}
\end{figure}

\paragraph{Rescaled correlations:}
To get a better understanding of the three states (a)-(c), we study the properties of the rescaled one-particle density matrix,
\begin{equation}
\label{eq:corrscal}
C_j(r)=\aver{b^\dagger_j b^{\phantom{\dagger}}_{j+r}}/\sqrt{n_j n_{j+r}},
\end{equation}
in which the leading density dependence of $b_j\propto \sqrt{n_j}$ is divided out. In the absence of density fluctuations $C_j(r)$ is just the pure phase correlation function $
\aver{\e^{{\mathrm i}\phi_j} \e^{-{\mathrm i}\phi_{j+r}}}$. At the two particle level, the equivalent step is going from the two-particle density $\rho^{\left(2\right)}(x_1,x_2)$ to the dimensionless two particle distribution function $g^{\left(2\right)}(x_1,x_2)= \rho^{\left(2\right)}(x_1,x_2)/\rho^{(1)} (x_1) \rho^{(1)}(x_2)$. 
Remarkably, we find that by this simple rescaling, the signatures of the SF and MI phases in the homogeneous system, namely an algebraic or exponential decay, $C_{j}(r)\propto A \norm{r}^{-K/2}$ and $\propto B\e^{- \norm{r}/\xi}$, respectively, can be recovered approximately even in the presence of a parabolic confining potential.
For weak interactions, $u\le u_{c1}$, [Fig.\ \ref{fig:corrdecaycomb} (a)] $C_j(r)$ decays approximately algebraically with $r$.% (neglecting the boundary regions).
 In the intermediate regime, $u_{c1}<u<u_{c2}$, [Fig.\ \ref{fig:corrdecaycomb} (b)] the decay in the regions where the density is incommensurate is still algebraic, whereas in the regions where the density is locked, it shows an exponential behaviour. Increasing the interaction further, $u\ge u_{c2}$, [Fig.\ \ref{fig:corrdecaycomb} (c)] the incommensurate regions disappear and the correlations decay exponentially. 

\begin{figure} 
\begin{center}
        {\epsfig{figure=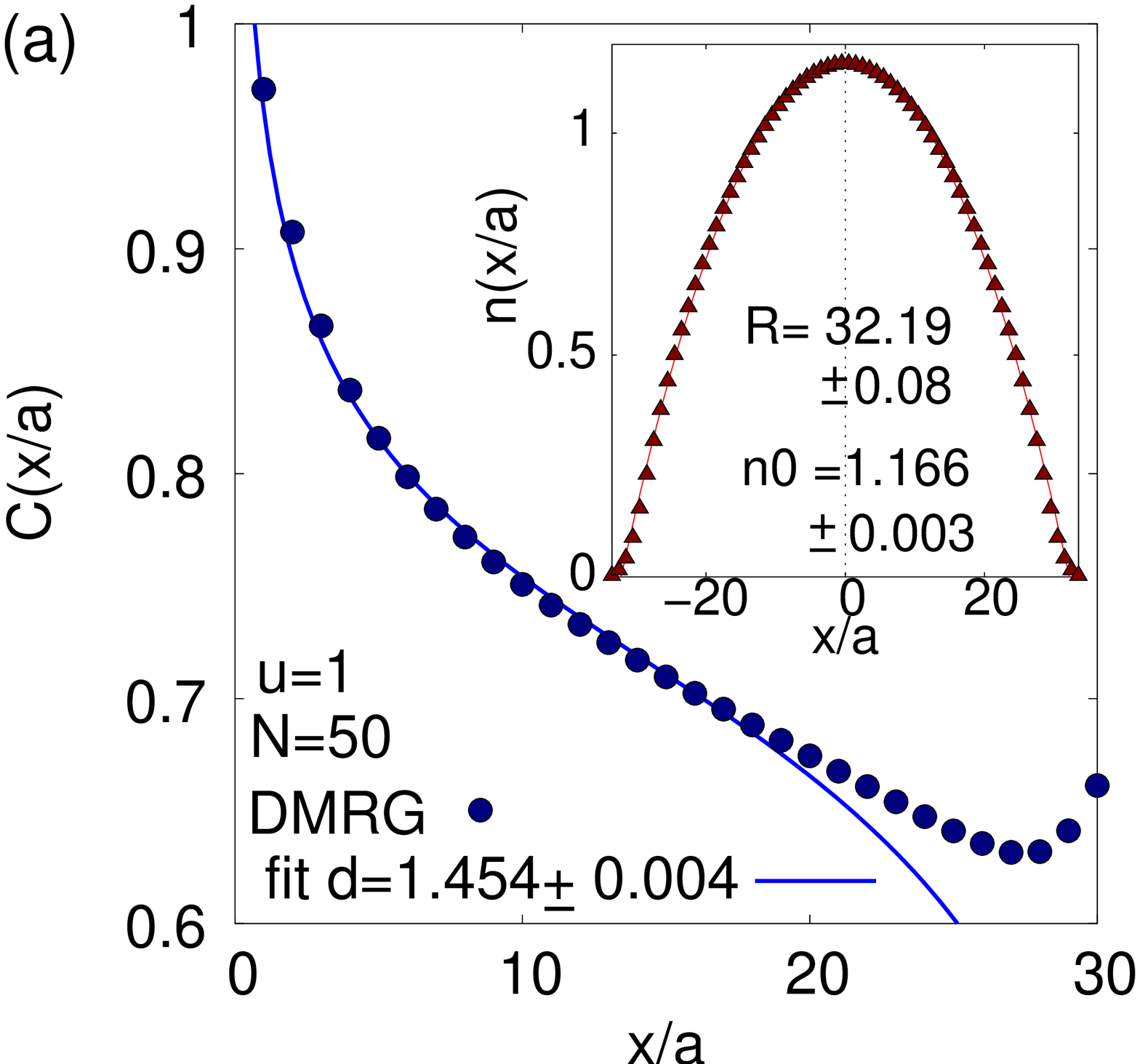,width=0.48\linewidth}}
        {\epsfig{figure=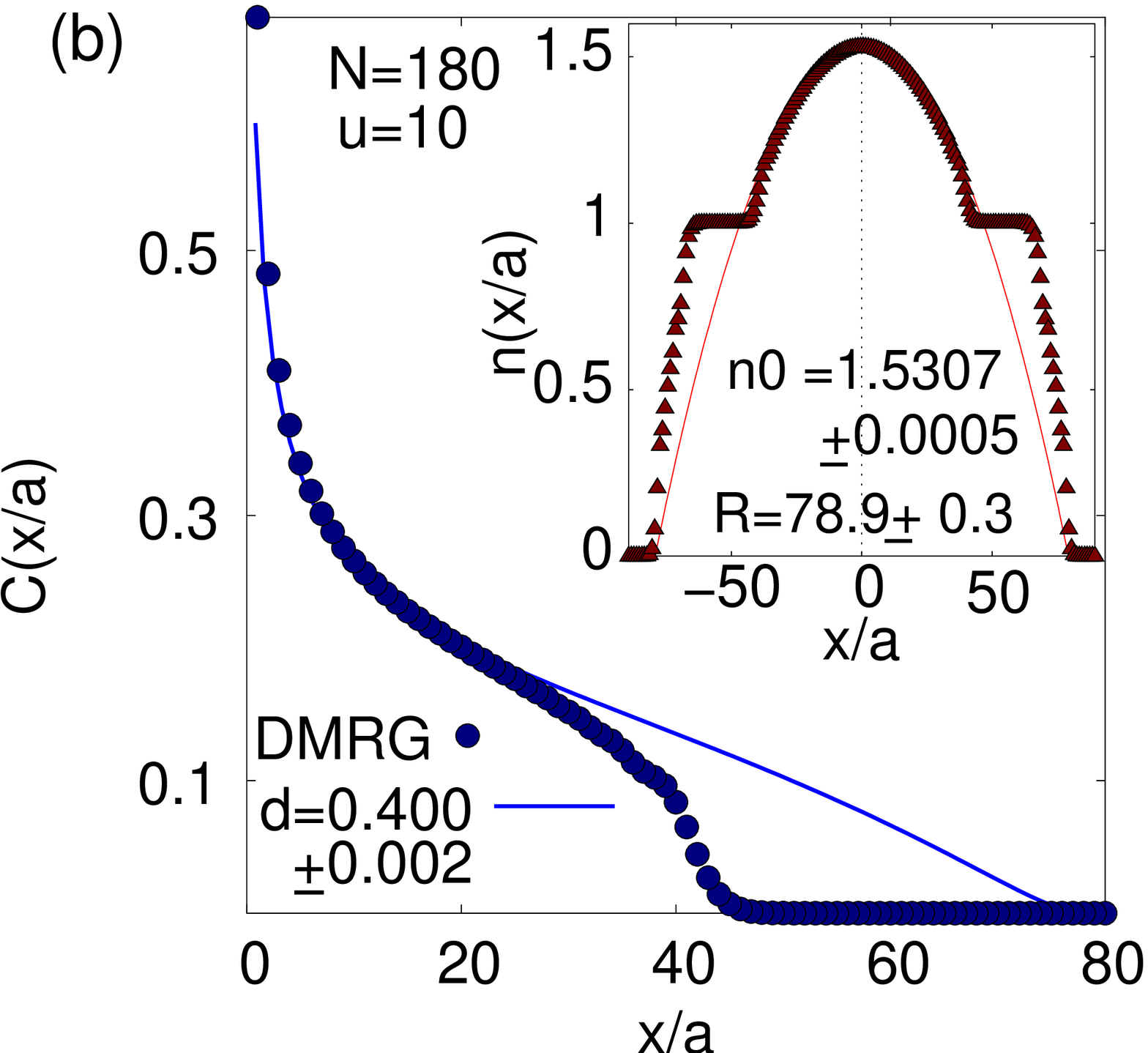,width=0.48\linewidth}}
\end{center}
\caption{Quasi-exact DMRG results for $C(j)$ (symbols) are compared to equation (\ref{eq:corrsym}) obtained by the hydrodynamical approach \cite{GangardtShlyapnikov2003} (lines). We used $n(x)=n_0 (1-(x/R)^2)$, where $n_0$ and $R$ are determined by fitting to the DMRG results (see insets). The uncertainties are obtained by varying the fit range in the sensible region away from the boundaries.  
}
\label{fig:corrsym}
\end{figure}

\paragraph{Hydrodynamical approach:}
It is instructive to compare the numerically exact DMRG results to a hydrodynamical treatment of the interacting 1D Bose gas \cite{Haldane1981} combined with a local density approximation.
In the hydrodynamical approach the low-energy fluctuations of the system are described by two conjugate fields, the phase fluctuations $\phi(x)$ and the density fluctuations $\theta(x)$. 
This approach can be generalized to the case of inhomogeneous systems \cite{GangardtShlyapnikov2003} by taking the density fluctuations around a smooth, spatially dependent density profile $n(x)$. An equivalent procedure was used for 1D Fermionic gases by Recati et al. \cite{RecatiZoller2003}. The Hamiltonian becomes
\begin{equation*}
H=\frac{\hbar}{2 \pi} \int dx \left\{ v_j(x) (\partial_x \phi)^2+v_N(x) [\partial_x \theta-\pi n(x)]^2\right\},
 \end{equation*}  
precisely as in the homogeneous case, except that $n(x)$, and therefore $v_j(x)=\pi \hbar n(x) /m$ and $v_N(x)=(\pi \hbar)^ {-1}(\frac{\partial \mu}{ \partial n})|_{n=n(x)}$, now depend on $x$. 
To account for the inhomogeneity, the local density approximation $\mu[n(x)]+V(x)=\mu[n(0)]$
was used to obtain the mean density profile \cite{DunjkoOlshanii2001}. Based on this approximation Gangardt and Shlyapnikov \cite{GangardtShlyapnikov2003} have shown that the normalized matrix elements of the one-particle density matrix are given by:
\begin{eqnarray}
&&C(x):= \frac{\aver{b^\dagger(x) b(-x)}} { \sqrt{n(x)n(-x)}}=\left(\frac{ \norm{2x}}{l_c(x)}\right)^{-K(x)/2}, \label{eq:corrsym}%\\
%%l_c(x)&&\approx \sqrt{\frac{d}{n(x)}},\; K(x)\approx \left( \pi \sqrt{d n(x)}\right)^{-1},
%\textrm{ and } d=\frac{\hbar^2}{mg}.\nonumber
\end{eqnarray}
%{\bf
 where $K$ is the exponent and $l_c$ the longitudinal correlation length. Eq. (\ref{eq:corrsym}) is derived assuming $\norm{2x}\gg l_c$. Specializing to weak interaction, i.e. $\gamma\equiv 1/d n\ll 1$, the approximations
$l_c(x)\approx \sqrt{\frac{d}{ n(x)}}$ and $K(x)\approx1/ \left( \pi \sqrt{d n(x)}\right)$ hold, where $d\propto l_{\perp}^2/a_{3D}$ is the characteristic length of the interaction. $d$ depends on the 3D scattering length $a_{3D}$ and the amplitude $l_{\perp}$ of the transverse zero point oscillation.
%\textrm{ and } d=\frac{\hbar^2}{mg}.\nonumber
The condition $\norm{2x}\gg l_c$ breaks down at the boundaries, where $n(x)$ vanishes causing a divergence in $l_c(x)$.
%}
Comparing [Eq.\ (\ref{eq:corrsym})] to the quasi-exact results of DMRG, it turns out that the local density approach describes very well the rescaled correlations in the inhomogeneous systems for $\gamma\leq 2$.
To this end we fitted the function $C(x)$ [Eq.\ (\ref{eq:corrsym})] to the corresponding DMRG results, using only $d$ as fitting parameter [Fig.\ \ref{fig:corrsym}].
%, extracting $\gamma$. 
We find very good agreement in the bulk of the SF regions in both, the purely SF state [Fig.\ \ref{fig:corrsym} (a)] and the coexistence state [Fig.\ \ref{fig:corrsym} (b)]. The quality of the agreement is somewhat surprising, because 
%the values obtained for 
the pure state ($\gamma=0.6$) and the coexistence state ($\gamma=1.7$) are in an intermediate regime between the Thomas-Fermi limit ($\gamma \ll 1$) and the Tonks gas ($\gamma \gg 1$), where the density profile is no longer parabolic \cite{DunjkoOlshanii2001}.
%to the parameter regimes specified in \cite{DunjkoOlshanii2001}, we find that our systems are not deep in the Thomas-Fermi regime ($\gamma \ll 1$), but in an intermediate regime between the latter and the Tonks gas ($\gamma \gg 1$).

\begin{figure} 
\begin{center}
        {\epsfig{figure=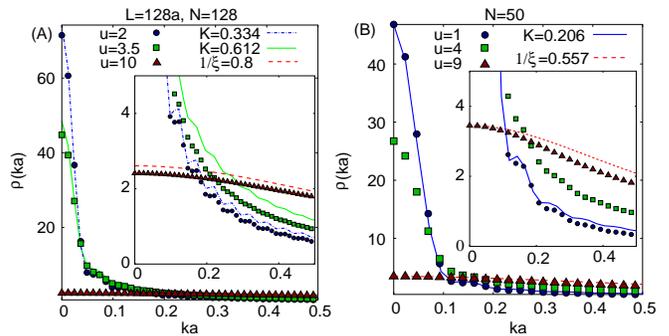,width=\linewidth}}
\end{center}
\caption{Interference pattern for the system with (A) open boundaries and with (B) parabolic trap for different values of $u$. Symbols are the results of the DMRG (maximal uncertainty $0.1$) and lines the results of the approximations explained in the text. The insets enlarge the scale of the $y$-axis. For a homogeneous system $u_c(n=1)\simeq 3.37$ is the critical value in the thermodynamic limit according to \cite{KuehnerMonien2000}.}
\label{fig:interference}
\end{figure}

\begin{figure} 
\begin{center}
        {\epsfig{figure=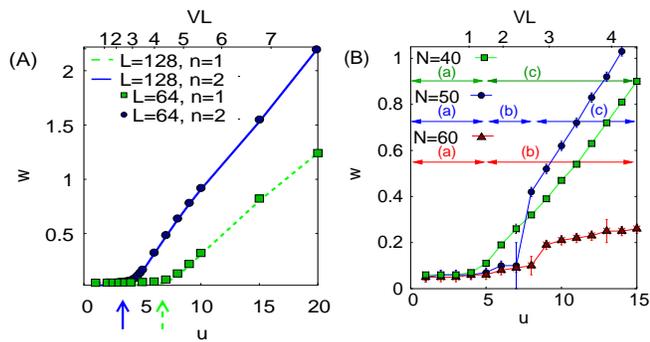,width=\linewidth}}
\end{center}
\caption{Half width of the interference peak for the homogeneous (A) and the parabolic (B) system. Arrows in (A) mark the critical value of $u_c$ in the thermodynamic limit (solid and dashed for $n=1$ and $n=2$ respectively) according to \cite{KuehnerMonien2000}. Arrows in (B) mark the three different regimes described in the text. To relate $u$ to the corresponding lattice depth $VL=V_{\textrm{lat}}/E_r$ of experiments, we assumed that the depths in the two perpendicular dimensions where fixed to $V_{\textrm{lat},\perp}/E_r=50$. } 
\label{fig:halfwidth}
\end{figure}

\emph{Interference pattern:}
We investigate how the information contained in the interference pattern is influenced by the confining potential.
%{\bf
If the interaction between the atoms after switching off the confining potentials is weak, i.e. $E_{\textrm{pot}}\ll E_{\textrm{kin}}$, the measured absorption images reflect the momentum distribution obtained from the Fourier transform of the one-particle density matrix \cite{KashurnikovSvistunov2002}
\begin{eqnarray}
I(k) \propto 
\rho(k)=\frac{1}{N}\sum_{j,j'=1}^{M} \e^{{\mathrm i} (j-j')ak} \aver{b^{\dagger}_{j} b^{\phantom{\dagger}}_{j'}},
\label{eq:interf}
\end{eqnarray}
where $M$ is the number of sites in the chain and $N$ the total number of particles. For the parameters studied here, the approximation of a negligible contribution of the interaction energy to the time of flight images is valid for all momenta in the second or in higher Brillouin-zones. Indeed, these momenta are of order $2\hbar \pi s /L$, where $s\in {\mathbbm N}$ and $s>M$. Thus $\frac{E_{\textrm{pot}}}{E_{\textrm{kin}}}\propto \frac{ n_{3D} (4 \pi \hbar^2 a_s)/m }{(\pi\hbar s /L)^2/(2m)}\propto\frac{a_s}{a}\propto 10^{-2} $ for $n_{3D}<1.5/a^3$ and $a_s/a$ like in \cite{GreinerBloch2002}. %}
The function $\rho(k)$ has been studied for very small systems numerically \cite{RothBurnett2002}, with the hydrodynamical approach \cite{Cazalilla2002} for a 1D homogeneous system and for the confined system in 3D \cite{KashurnikovSvistunov2002} and 1D \cite{PupilloWilliams2003}.
In Fig.\ \ref{fig:interference} we plot the DMRG results (symbols) for the function $\rho(k)$ for several values of the parameter $u$, comparing the homogeneous system ($\varepsilon_i=0$) with open boundary conditions (A) to the parabolic system (B).
In the homogeneous system with commensurate filling, $n=N/M$=1 [Fig.\ \ref{fig:interference} (A)] we find a very sharp peak at small momenta for $u<u_c$. If $u$ is increased, the peak height decreases smoothly. 
The half width $w$ [Fig.\ \ref{fig:halfwidth} (A)], however, shows a clear upturn. This upturn signifies a phase transition, since it stems from the behaviour of the correlation length $\xi$ ($\propto w^{-1}$), which diverges in the SF phase ($\xi\propto L$) and becomes finite in the MI phase ($\xi\propto \Delta^{-1}$, where $\Delta$ is the energy gap).
For the parabolic system [Fig.\ \ref{fig:interference} (B)], the interference pattern for small and large $u$ is  similar to the interference pattern in the homogeneous system. In the intermediate regime, however, it shows a more complex behaviour, which is most clearly evident in $w$ [Fig.\ \ref{fig:halfwidth} (B)].
For small particle numbers ($N=40$), $w$ is very small for $u\lesssim u_{c1}$ and rises continuously for $u> u_{c1}$. 
In contrast, for larger particle numbers ($N=50,60$) three different regimes corresponding to the three different states in Fig.\ \ref{fig:statediasketch} are observed:
(a) for $u<u_{c1}$, $w$ is very small, (b) for  $u_{c1}<u<u_{c2}$, $w$ rises slowly, until at $u\sim u_{c2}$ it shows a sudden jump-like increase, (c) for $u>u_{c2}$, it continues to rise strongly.  
That means that in the SF (a) and the MI (c) state the behaviour of $w$ resembles that of the homogeneous system. This is as expected, since the rescaled correlations show the same decay as in the corresponding homogeneous phases. In the intermediate regime (b), however, it shows a new behaviour, a slow increase, which is due to the coexistence of the SF and the MI state. The SF region determines mainly the height of the interference peak, while its broadening is due to the presence of the MI region. In the crossover region between the totally incommensurate and the coexistence region, the interference pattern shows additional ocillations with period $2\pi/l$, where $l$ is the distance between the two outer SF regions, due to the appearance of relatively strong correlations between the latter. Similar oscillations were seen in \cite{PupilloWilliams2003}. In smaller systems such as in \cite{KashurnikovSvistunov2002} the effect is more pronounced causing well-separated satellite peaks . 
%However we did not find single side peaks in contrast to the calculations in three dimensions in \cite{KashurnikovSvistunov2002}.

Finally, let us investigate to what extent the properties of the interference patterns in Fig.\ \ref{fig:interference} can be understood in terms of simple phenomenological approximations for $\aver{b^{\dagger}_{j} b^{\phantom{\dagger}}_{j'}}$ in the homogeneous and the rescaled correlations $C_j(r)$ in the inhomogeneous system. Once the characteristic quantities $K$ and $\xi$ have been identified (in this case by fitting to DMRG results), our simple rescaling procedure captures most of the essential observable physics. To illustrate this we show in Fig.\ \ref{fig:interference} (A) in addition to the DMRG results, results (lines) obtained by approximating $\aver{b^{\dagger}_{j} b^{\phantom{\dagger}}_{j'}}$ in Eq.\ (\ref{eq:interf}) by $ A\norm{j-j'}^{-K/2}$ and $ B\e^{-\norm{j-j'}/\xi}$ for small and large $u$, respectively.
The values of $K$ and $\xi$ are determined by fitting $\aver{b^{\dagger}_{j} b^{\phantom{\dagger}}_{j'}}$ to DMRG results (not shown here). The constants $A$ and $B$ are chosen such that the value at $k=0$ agrees with the DMRG results. In Fig.\ \ref{fig:interference} (B) the approximation (lines) are obtained analogously by taking the density scaling into account, i.e. replacing $\aver{b^{\dagger}_{j} b^{\phantom{\dagger}}_{j'}}$ by the algebraically and the exponentially decaying functions times the scaling factor $\sqrt{n_j n_{j'}}$. We use the density distribution $n_j=n_0 (1-(j-j_0)^2/R^2)$ for $u=1$, and $n_j=1$ for $u=9$. The parameters $K$ and $\xi$ are determined by fitting the rescaled correlation functions. Comparing the DMRG data to the approximation we see in Fig.\ \ref{fig:interference} that this simple approximation works very well for small values of $ka$; in particular, it reproduces the correct shape of the peak [even including the small non-monotonities which are due to the finite sum in Eq.\ (\ref{eq:interf})]. This underlines that $\rho(k)$ is mainly determined by the decay of the (un)scaled correlations.
Clearly our calculations in 1D cannot be compared quantitatively with the experiments in a 3D lattice \cite{GreinerBloch2002}. Recently, however, an array of truly 1D Bose systems has been created \cite{MoritzEsslinger2003}. With an additional lattice potential our predictions can then be tested quantitatively \cite{StoeferleEsslinger2003}. 
%one may then study our predictions. 
In the experimental realization one typically has several 1D systems next to each other with different particle number, hence the location of the sharp upturn in the half width [Fig.\ \ref{fig:halfwidth} (B)], will be smeared out, since the critical value $u_{c2}$ depends on the particle number. Nevertheless, we expect in particular the strong, jump-like increase between the coexistence state and the MI state to remain observable.

In conclusion, we have found that the correlation functions of a parabolically confined system, after a remarkably simple rescaling, show approximately the familiar algebraic and exponential behaviour of the SF and MI phases in the homogeneous system. We investigated as well the applicability of the local density approximation in a parabolic system in the limit of weak interaction and find good agreement with the DMRG results. Moreover, if the experimental system consists of 1D tubes with almost the same average filling, the half width of the interference peak can be used to distinguish the different types of states that occur experimentally.
\acknowledgments

We would like to thank M. Cazalilla, I. Bloch, M. Greiner, I. Cirac, J.J. Garcia-Ripoll, and T. Giamarchi for fruitful discussions. C. Kollath was financially supported by the Hess-Preis and project DE 730/3-1 of the DFG and the Studienstiftung des deutschen Volkes.

%\bibliography{references}

\end{document}